# A novel paradigm for solving PDEs: multi scale neural computing

Wei Suo[1], Weiwei Zhang[1]

1 Northwestern Polytechnical University, 710072 Xi' an, People' s Republic of China

**Abstract**

Numerical simulation is dominant in solving partial difference equations (PDEs), but balancing fine-grained grids with low computational costs is challenging. Recently, solving PDEs with neural networks (NNs) has gained interest, yet cost-effectiveness and high accuracy remains a challenge. This work introduces a novel paradigm for solving PDEs, called multi scale neural computing (MSNC), considering spectral bias of NNs and local approximation properties in the finite difference method (FDM). The MSNC decomposes the solution with a NN for efficient capture of global scale and the FDM for detailed description of local scale, aiming to balance costs and accuracy. Demonstrated advantages include higher accuracy (10 times for 1D PDEs, 20 times for 2D PDEs) and lower costs (4 times for 1D PDEs, 16 times for 2D PDEs) than the standard FDM. The MSNC also exhibits stable convergence and rigorous boundary condition satisfaction, showcasing the potential for hybrid of NN and numerical method.

**Keywords**: Neural computing, Partial difference equations, Hybrid strategy, Numerical methods, Neural networks

## 1 Introduction

Numerical simulation is a dominant approach for solving partial difference equations (PDEs) in scientific research and engineering application. However, high-fidelity simulations depend on fine-grained spatial and temporal discretization to resolve all phenomena of interest. It leads to inevitable trade-offs between computational costs and accuracy, becoming a bottleneck of numerical methods.

In recent years, neural networks (NNs) has attracted increasing interest in solving PDEs[1-11]. Physics-informed neural networks (PINNs)[1] incorporate the governing equations of the system into the loss function. By optimizing the NNs parameters,

PINNs gradually approximate the governing equations and boundary conditions, enabling the solution of partial differential equations (PDEs). Another type of NNs, represented by extreme learning machine (ELM), involves solving for the output layer weights only, while the parameters of other neurons are randomly generated. ELM establishes algebraic systems of equations for output layer weights based on PDEs and boundary conditions, and then solves them[9-11]. Despite the advantages of NNs-based PDEs solvers in inverse problems and no explicit need for a mesh generation, they encounter the limitation in optimizing high-dimensional non-convex functions. The current NNs-based solvers struggle to match the efficiency and accuracy of numerical methods in solving forward problems[12-13].

Hybrid of numerical methods and NNs, by leveraging the inference capabilities of NNs to enhance numerical methods, aiming to improve the accuracy and reduce computational costs in solving PDEs, receives increasing attention. The existing hybrid strategies can be fallen into several categories, including data-driven shock capturing methods, data-driven correction of numerical schemes, and data-driven discretizations. Data-driven shock capturing methods utilize NNs to enhance the capability of shock-capturing of numerical methods[14-20]. Discacciati et al.[14] designed artificial neural networks (ANNs) to predict the artificial viscosity, guaranteeing optimal accuracy for smooth problems, as well as good shock-capturing properties. Stevens et al.[20] trained a neural network to modify weights of WENO5-JS, thereby improving the accuracy. Data-driven correction of numerical schemes utilize NNs to predict error discrepancy between low-order and high-order schemes, in order to enhance the accuracy of low-order schemes[21-24]. de Lara and Ferrer [21-22] proposed a method to accelerate high-order Discontinuous Galerkin (DG) methods using NNs. NNs are trained to extract a corrective forcing of lower order solution for recovering high order accuracy. Kossaczká et al.[24] introduced a deep finite difference method (DFDMs) based on an approximation of the local truncation error of the finite difference method (FDM) by convolutional neural networks (CNNs). Data-driven discretizations utilize NNs to enhance the discretization accuracy of spatial derivatives terms[25-28]. Bar-Sinai et al.[25] employed CNNs to learn low-resolution discrete models that encapsulate unresolved physics, improving the accuracy and stability of simulations on coarse grids. Kiyani et al.[28] used solution and numerical differentials from the previous time step as inputs to a multilayer perceptron (MLP), with the advancement of the current time step as outputs.

The existing hybrid strategies have demonstrated the feasibility of enchancing numerical methods by NNs. However, the current strategies are established without taking into account the spectral bias of NNs and local approximation properties of numerical methods. NNs exhibit a spectral bias towards fitting smooth functions, with better capabilities in fitting low-frequency and global information[29]. In numerical methods such as the FDM, the solution domain is discretized into multiple local regions, and polynomials are employed to approximate segments of solution in these local regions. In response to these considerations, we propose a novel paradigm for solving PDEs, called multi scale neural computing (MSNC). The MSNC is based on the idea of scale decomposition, consisting of a NN for efficient capture of global scale and the FDM for detailed description of local scale. The motivation of the MSNC is to balance the computational costs and accuracy, making the solution of PDEs more convenient.

## 2 Results

### 2.1 Hybrid Strategy of NNs and the FDM

First, let us focus on the hybrid strategy of numerical methods and NNs. Xu et al.[29] revealed the spectral bias of NNs, stating that NNs tend to learn low-frequency information first when adapting to new signals, and then slowly learn high-frequency information. The spectral bias inspires the hybrid strategy in this paper. For a globally defined NN, if the spatial coordinates $x$ are taken as input and the solution of PDEs $u$ as output, it may tend to fit low-frequency global smooth information and may have insufficient fitting for local information during the training. But for numerical methods like the FDM, since trial functions locally approximate the solution, local errors decay faster during the solving process, while the decay of global offset errors is slower, manifesting as low-frequency errors. It is evident that NNs and numerical methods can complement each other. For NNs, the introduction of numerical methods can eliminate local errors. For numerical methods, the introduction of neural networks can correct the global offset. Taking into account the the spectral bias of NNs and the local approximation properties of numerical methods, we propose a novel hybrid strategy of numerical methods and NNs as follows. The solution of PDEs are approximated by combination of global solution and local solution, introduced by NNs and numerical methods respectively. Based on the idea of scale decomposition, NNs are employed for efficient capture of global scale, while numerical methods are utilized for detailed description of local scale. Given the combination of global and local scale, we name the proposed hybrid strategy as multi scale neural computing.

Then we need to address the question of how NNs are generated in the MSNC. NNs have the potential to leverage historical data, process data, and experimental data to reduce the computational costs of numerical simulations. Therefore, we adopt the following data-driven NNs generation strategy. Firstly, train the NNs using historical data that is similar to the state to be solved. Leveraging the similarity among data, the obtained NNs can better approximate the solution compared to random generation. Then, drawing inspiration from the ELM concept, we release the output layer weights of the trained NNs as free parameters in the new state. This idea of releasing certain degrees of freedom (DOFs) in the NNs is akin to the fine-tuning concept in transfer learning[30]. Finally, during the solution of the new state, similar to the ELM, the output layer weights are unknowns to be solved and the NNs is then combined with numerical methods to solve the PDEs in the new state. Figure 1(a) illustrates this data-driven NNs generation strategy.

Finally, we illustrate the procedure of the MSNC as follows. The solution process can be divided into two stages: offline and online. In the offline stage, the NN is trained based on historical data, and its output layer weights are released. In the online stage, the NN and FDM are used to discretize the PDEs to be solved, with decomposing the variable $u$ into two parts, $u = \bar{u} + \hat{u}$, where $\bar{u}$ denotes the solution at the global scale, solved by the NN, and $\hat{u}$ denotes the solution at the local scale, solved by the FDM. For solution at the global scale, automatic difference (AD) of the NN is used to compute the derivatives. For solution at the local scale, the FDM such as the second-order central difference is applied to compute the derivatives. Due to the unknowns of NN are output layers weights, after discretization, an algebraic system of equations is formed, with unknowns including the output layer weights of the NN and the local values at each grid point in FDM. The overall solution process of the PDEs using the MSNC is illustrated in Figure 1(b).

a)

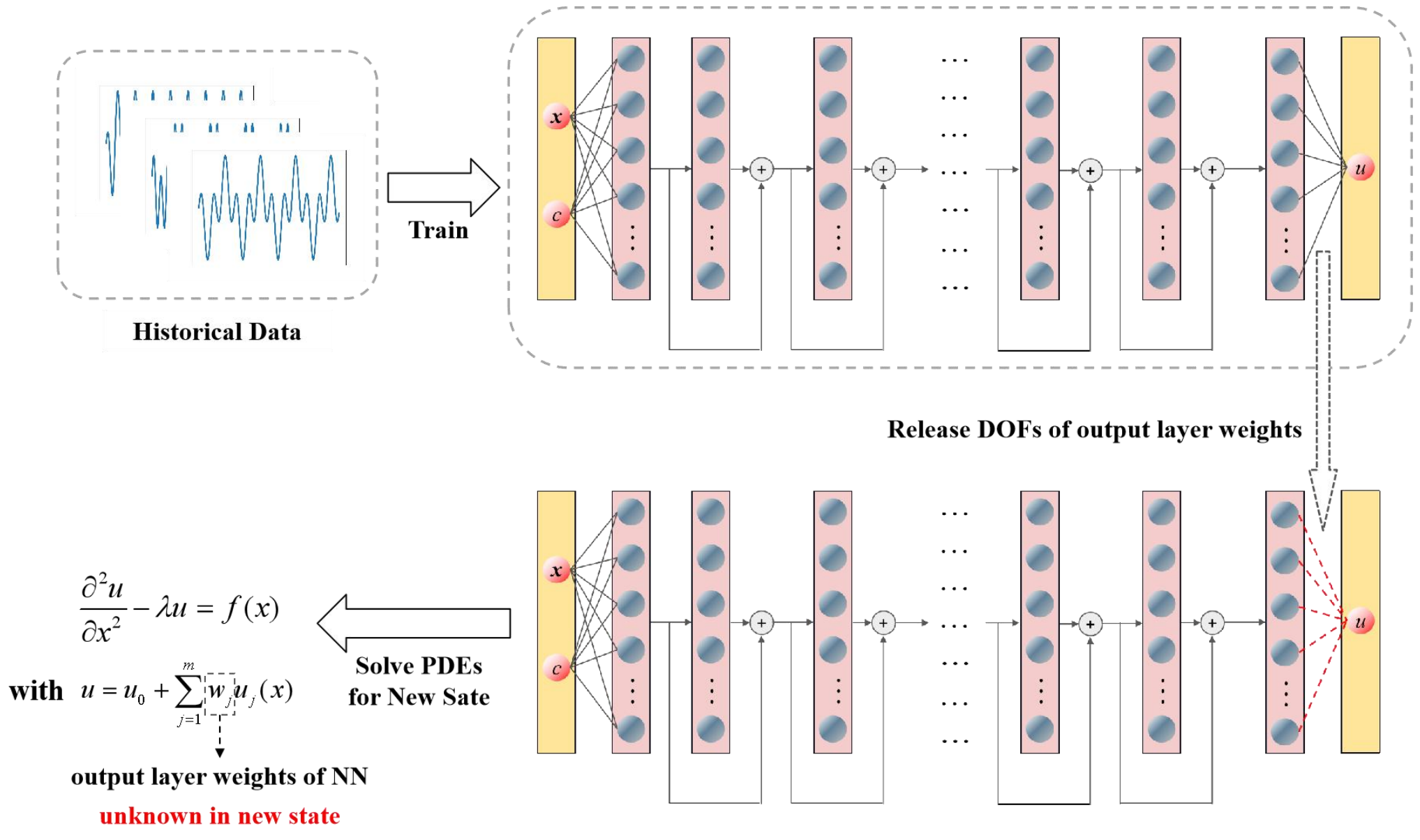

b)

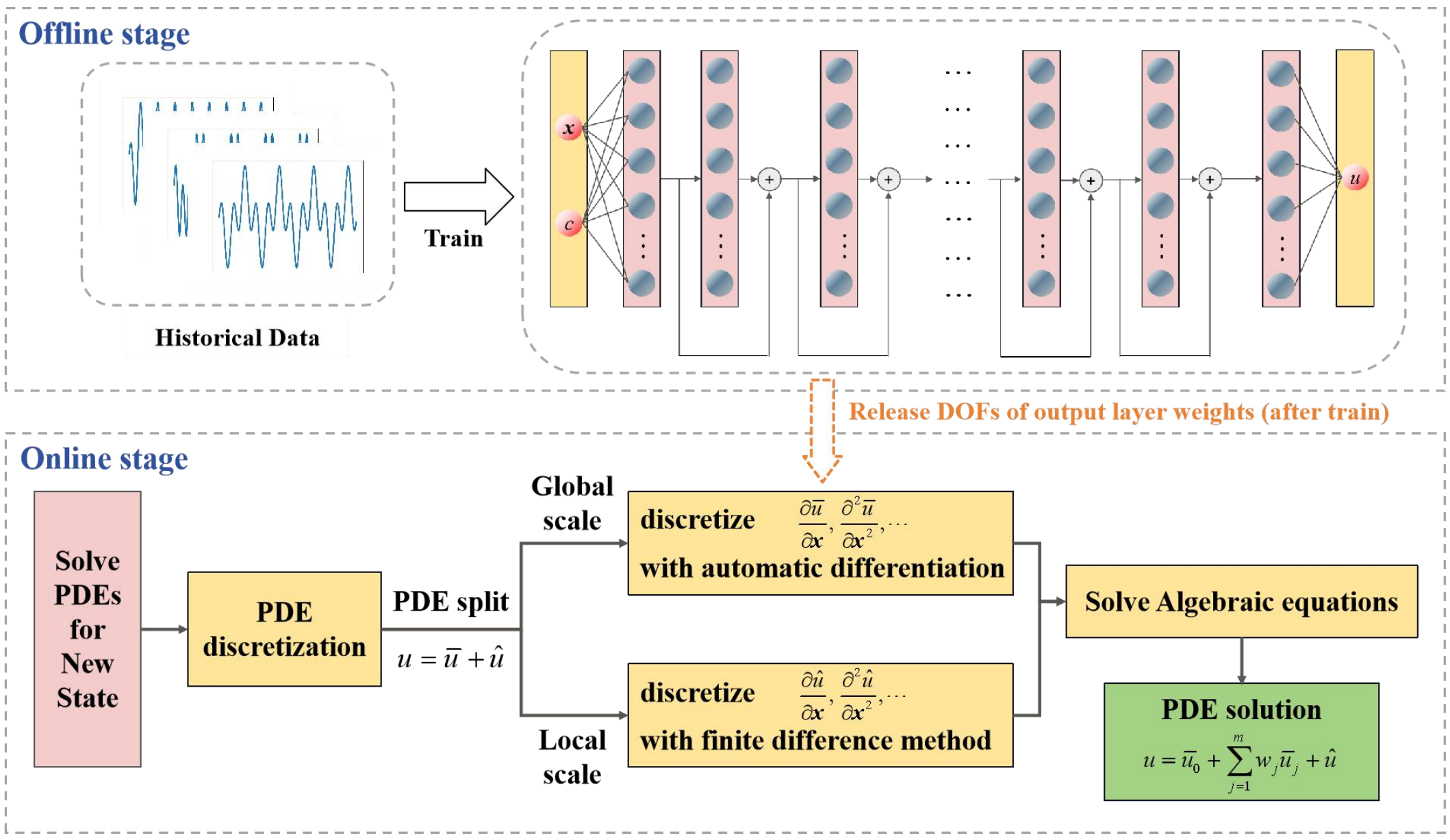

Figure 1: **a)** Data-driven neural network generation strategy. **b)** The solution process of the PDEs using the MSNC.

2.2 Solutions of one-dimensional PDEs

We consider the one-dimensional linear Helmholtz equation with smooth solution, the one-dimensional Poisson equation with steep solution, and the one-dimensional nonlinear Helmholtz equation with smooth solution.

2.2.1 Examples description

(I) One-dimensional linear Helmholtz equation (with smooth solution)

Consider the one-dimensional linear Helmholtz equation of the form

$$\begin{aligned} &\frac{\partial^2 u}{\partial x^2} - \lambda u = f(x), \quad x \in [a,b] \\ &u(a) = g_1, \quad u(b) = g_2 \end{aligned} \tag{1}$$

The constant and the domain specification in the Eq. (1) are $a = -4, \quad b = 4, \quad \lambda = 10$.

We choose the source term $f(x)$ such that the Eq. (1) has the following solution

$$u(x) = \sin(\alpha_1 \pi x)\cos(\alpha_2 \pi x) \tag{2}$$

where $\alpha_1$ and $\alpha_2$ are two state parameters used for sampling. For given state parameters $\alpha_1$ and $\alpha_2$, $g_1$ and $g_2$ are chosen according to analytic solution by setting $x = a$ and $x = b$ in Eq. (2), respectively. Additionally , we choose $\alpha_1 = 3$ and $\alpha_2 = 2$ as the test state for this example.

(II) One-dimensional Poisson equation (with steep solution)

Consider the one-dimensional Poisson equation of the form

$$\begin{aligned} &\frac{\partial^2 u}{\partial x^2} = f(x), \quad x \in [a,b] \\ &u(a) = g_1, \quad u(b) = g_2 \end{aligned} \tag{3}$$

The domain specification in the Eq. (3) are $a = -4, \quad b = 4$.

We choose the source term $f(x)$ such that the Eq. (3) has the following solution

$$u(x) = 0.5\sin(\alpha_1 \pi x)\cos(\alpha_2 \pi x) + \tanh(20x) \tag{4}$$

where $\alpha_1$ and $\alpha_2$ are two state parameters used for sampling. For given state parameters $\alpha_1$ and $\alpha_2$, $g_1$ and $g_2$ are chosen according to analytic solution by setting $x=a$ and $x=b$ in Eq. (4), respectively. $\tanh(20x)$ in Eq. (4) is used for simulating steep gradients to test the accuracy of the MSNC in solving 1-D PDEs with steep solution. Additionally , we choose $\alpha_1=3$ and $\alpha_2=2$ as the test state for this example.

(III) One-dimensional nonlinear Helmholtz equation (with smooth solution)

Consider the one-dimensional Poisson equation of the form

$$\begin{aligned} &\frac{\partial^2 u}{\partial x^2}=f(x), \quad x\in[a,b] \\ &u(a)=g_1, \quad u(b)=g_2 \end{aligned} \tag{5}$$

The domain specification in the Eq. (5) are $a=-4, \quad b=4$.

We choose the source term $f(x)$ such that the Eq. (5) has the following solution

$$u(x)=0.5\sin(\alpha_1\pi x)\cos(\alpha_2\pi x)+\tanh(20x) \tag{6}$$

where $\alpha_1$ and $\alpha_2$ are two state parameters used for sampling. For given state parameters $\alpha_1$ and $\alpha_2$ , $g_1$ and $g_2$ are chosen according to analytic solution by setting $x=a$ and $x=b$ in Eq. (6), respectively. $\tanh(20x)$ in Eq. (6) is used for simulating steep gradients to test the accuracy of the MSNC in solving 1-D PDEs with steep solution. Additionally , we choose $\alpha_1=3$ and $\alpha_2=2$ as the test state for this example.

2.2.2 Numerical results

We introduce a uniform grid to discretize the 1D spatial domain $[a,b]$. The grid numbers is denoted by *gridNum*. In addition, the standard FDM is employed as a comparison to the MSNC.

Figure 2 show the solving results, including comparison of solution, point-wise error, and decomposition of the MSNC, of the one-dimensional PDEs with *gridNum* of 320. We can observe that, the MSNC achieves higher accuracy compared to the standard FDM, with the magnitude of the point-wise error mainly between $O(10^{-4})$ and $O(10^{-2})$ in the standard FDM ,while mainly between $O(10^{-5})$ and $O(10^{-3})$ in the MSNC. Moreover, both the MSNC and standard FDM exhibit minimal errors at the boundary points, indicating the boundary conditions handling ensure rigorous satisfaction of boundary conditions. Additionally, in the decomposition of the MSNC, we can see that the NN solution provides a global profile similar to the entire solution, while the FDM solution offers local refinement to capture more local details. This demonstrates the feasibility of the MSNC ’ s starting point, where NNs are employed for efficient capture of global scale, while the FDM are utilized for detailed description of local scale.

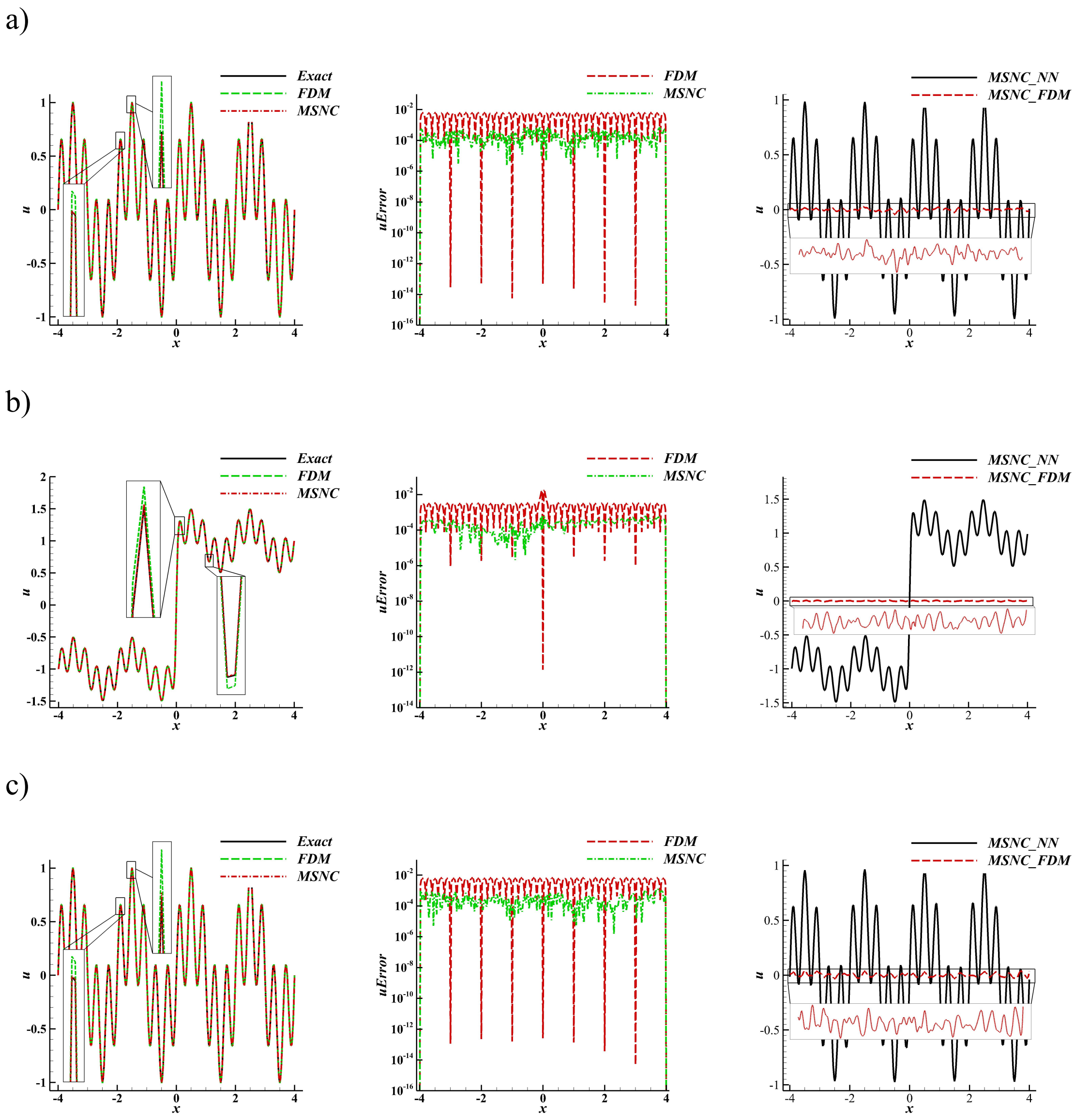

Figure 2: **a)** One-dimensional linear Helmholtz equation: comparison of solution, point-wise error, and decomposition of the MSNC (*gridNum* = 320). **b)** One-dimensional Poisson equation: comparison of solution, point-wise error, and decomposition of the MSNC (*gridNum* = 320). **c)** One-dimensional nonlinear Helmholtz equation: comparison of solution, point-wise error, and decomposition of the MSNC (*gridNum* = 320).

Table 1 shows the RMSE with different *gridNum* and convergence order of the one-dimensional PDEs. We can observe that, with an equivalent grid number, the MSNC consistently achieves higher solution accuracy than the standard FDM. We compute the ratio, defined as the RMSE of the standard FDM divided by the RMSE of the MSNC at same *gridNum*. We can see that, except for the minimum value of *gridNum*, the ratio is generally more than 10, demonstrating the significant improvement in accuracy achieved by the MSNC compared to the standard FDM. From another perspective, at the same level of accuracy, the MSNC requires approximately 4 times fewer grid numbers compared to the standard FDM, effectively reducing computational costs. Moreover, we can observe that, after *gridNum* exceeding a certain point, the MSNC exhibits a approximate 2-order convergence similar to the standard FDM.

In summary, the above examples of one-dimensional PDEs demonstrates the MSNC 's notable advantages in both computational accuracy (exceeding a 10 times improvement) and cost (accompanied by a 4 times reduction in grid numbers) compared to the standard FDM.

## 2.3 Solutions of two-dimensional PDEs

We consider the two-dimensional Poisson equation with smooth solution and steep solution.

### 2.3.1 Examples description

Consider the two-dimensional Poisson equation of the form

$$\begin{aligned}&\frac{\partial^2 u}{\partial x^2}+\frac{\partial^2 u}{\partial y^2}=f(x,y),\quad x,y\in[a_1,b_1]\times[a_2,b_2]\\&u(a_1,y)=g_1,\quad u(b_1,y)=g_2,\quad u(x,a_2)=g_3,\quad u(x,b_2)=g_4\end{aligned}\tag{7}$$

The domain specification in the Eq. (35) are $a_1=-1,\quad b_1=1,\quad a_2=-1,\quad b_2=1$.

(I) Smooth solution

Table 1  One-dimensional PDEs: RMSE with different *gridNum* and convergence order

| | *gridNum* | RMSE FDM | Order | RMSE MSNC | Order | Ratio |
|---|---|---|---|---|---|---|
| One-dimensional linear Helmholtz equation | 80 | 7.821701e-02 | | 8.703153e-03 | | 8.99 |
| | 160 | 1.792897e-02 | 2.18 | 1.268052e-03 | 3.43 | 14.14 |
| | 320 | 4.392288e-03 | 2.04 | 2.744836e-04 | 2.31 | 16.00 |
| | 640 | 1.092990e-03 | 2.01 | 6.598891e-05 | 2.08 | 16.56 |
| | 1280 | 2.729848e-04 | 2.00 | 1.647682e-05 | 2.00 | 16.57 |
| | 2560 | 6.823646e-05 | 2.00 | 4.408690e-06 | 1.87 | 15.48 |
| One-dimensional Poisson equation | 160 | 1.380952e-02 | | 1.139017e-02 | | 1.21 |
| | 320 | 3.064816e-03 | 2.25 | 2.733005e-04 | 20.84 | 11.21 |
| | 640 | 7.569434e-04 | 2.02 | 6.210349e-05 | 2.20 | 12.19 |
| | 1280 | 1.887124e-04 | 2.01 | 1.814521e-05 | 1.71 | 10.40 |
| | 2560 | 4.715047e-05 | 2.00 | 4.243519e-06 | 2.14 | 11.11 |
| One-dimensional nonlinear Helmholtz equation | 80 | 8.159348e-02 | | 1.240237e-02 | | 6.58 |
| | 160 | 1.862403e-02 | 2.19 | 1.860972e-03 | 3.33 | 10.01 |
| | 320 | 4.556188e-03 | 2.04 | 3.859985e-04 | 2.41 | 11.80 |
| | 640 | 1.133388e-03 | 2.01 | 8.757908e-05 | 2.20 | 12.94 |
| | 1280 | 2.830505e-04 | 2.00 | 2.111212e-05 | 2.07 | 13.41 |
| | 2560 | 7.075104e-05 | 2.00 | 5.420050e-06 | 1.95 | 13.05 |

For smooth solution, we choose the source term $f(x)$ such that the Eq. (7) has the following solution

$$u(x, y) = \sin(\alpha_1 \pi x)\sin(\alpha_2 \pi y) \tag{8}$$

where $\alpha_1$ and $\alpha_2$ are two state parameters used for sampling. For given state parameters $\alpha_1$ and $\alpha_2$, $g_1$, $g_2$, $g_3$, and $g_4$ are chosen according to analytic solution at the left, right, lower, and upper boundaries in Eq. (8), respectively.

Additionally , we choose $\alpha_1 = 1$ and $\alpha_2 = 4$ as the test state for this example.

(II) Steep solution

For steep solution, we choose the source term $f(x)$ such that the Eq. (7) has the following solution

$$u(x, y) = (0.1\sin(\alpha_1 \pi x) + \tanh(10x)) \times \sin(\alpha_2 \pi y) \tag{9}$$

where $\alpha_1$ and $\alpha_2$ are two state parameters used for sampling. For given state parameters $\alpha_1$ and $\alpha_2$ , $g_1$ , $g_2$ , $g_3$ , and $g_4$ are chosen according to analytic solution at the left, right, lower, and upper boundaries in Eq. (9), respectively. $\tanh(10x)$ in Eq. (9) is used for simulating steep gradients to test the accuracy of the MSNC in solving 2-D PDEs with steep solution. Additionally , we choose $\alpha_1 = 2$ and $\alpha_2 = 2$ as the test state for this example.

2.3.2 Numerical results

We introduce a uniform grid to discretize the 2D spatial domain $[a_1, b_1] \times [a_2, b_2]$ . The grid numbers of single dimension is denoted by *gridNum1D*, with the corresponding grid numbers of two-dimensional domain denoted by *gridNum2D*. In addition, the standard FDM is employed as a comparison to the MSNC.

Figure 3 show the solving results, including comparison of solution, point-wise error, and decomposition of the MSNC, of the two-dimensional PDEs with *gridNum1D* of 80. We can see that the conclusions for one-dimensional PDEs can be extended to two-dimensional PDEs here. The MSNC exhibits higher accuracy than the standard FDM. Moreover, the feasibility of the MSNC 's starting point, where NNs are employed for efficient capture of global scale, while the FDM are utilized for detailed description of local scale, is demonstrated again in two-dimensional PDEs.

Table 2 shows the RMSE with different *gridNum1D* and convergence order of the two-dimensional Poisson equation. We can see that the conclusions for one-dimensional PDEs can be extended to two-dimensional PDEs here. The MSNC achieves higher solution accuracy than the standard FDM. We compute the ratio, which is generally more than 20, with the maximum value being 32.11 for smooth solution and 49.39 for steep solution, demonstrating the significant improvement in accuracy achieved by the MSNC. From another perspective, to achieve same level of accuracy, the MSNC requires approximately 4 times fewer grid numbers in single dimension, meaning 16 times fewer grid numbers in two-dimensional spatial domain. Moreover, we can observe that, The MSNC exhibits a approximate 2-order convergence similar to the standard FDM after *gridNum1D* exceeding a certain point.

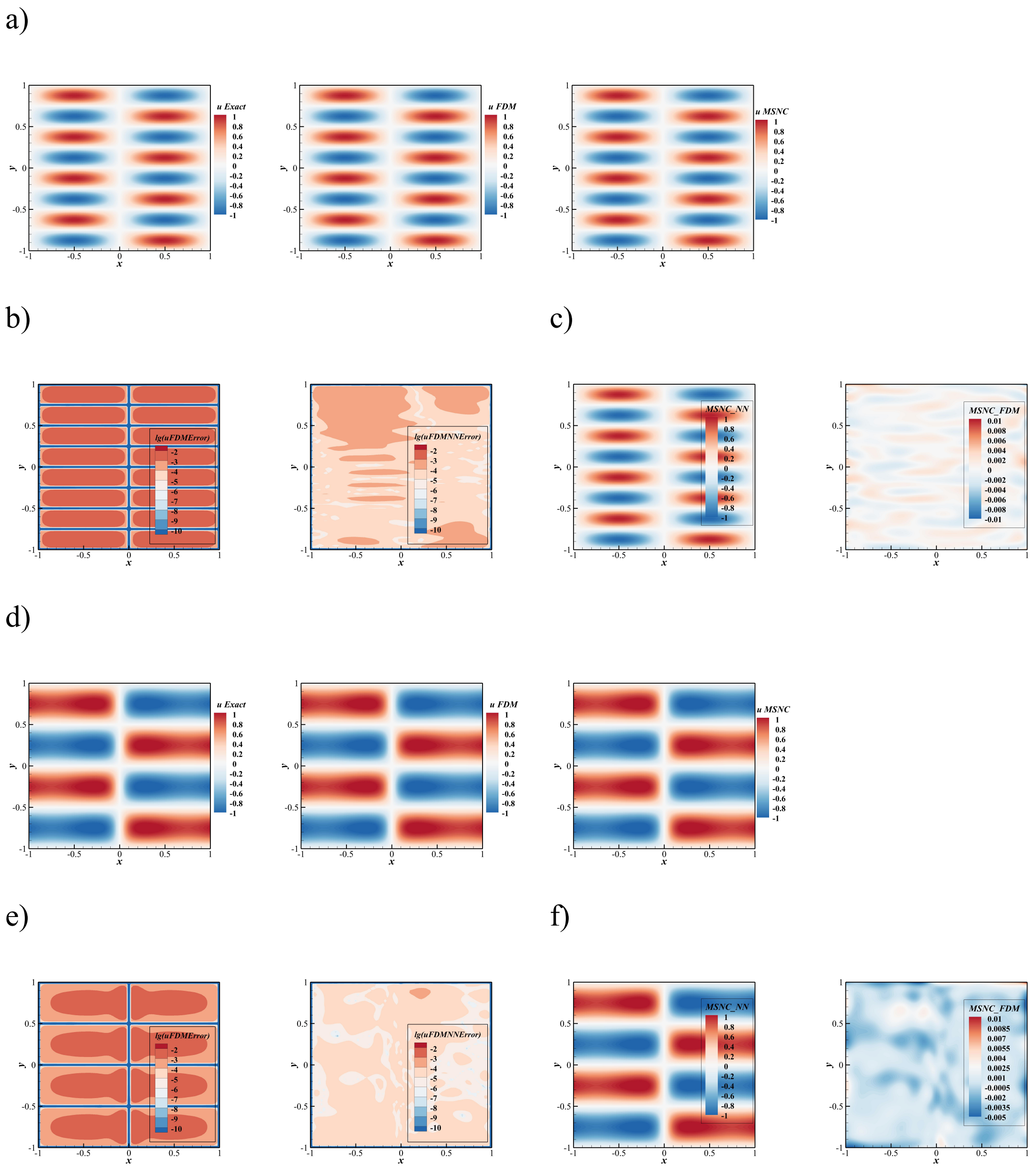

Figure 3: **a)** Two-dimensional Poisson equation with smooth solution: exact solution, FDM solution, and MSNC solution (*gridNum1D* = 80). **b)** Two-dimensional Poisson equation with smooth solution: point-wise error of the FDM and MSNC (*gridNum1D* = 80). **c)** Two-dimensional Poisson equation with smooth solution: decomposition of the MSNC with the NN and FDM part (*gridNum1D* = 80). **d)** Two-dimensional Poisson equation with steep solution: exact solution, FDM solution, and MSNC solution (*gridNum1D* = 80). **e)** Two-dimensional Poisson equation with steep solution: point-wise error of the FDM and MSNC (*gridNum1D* = 80). **f)** Two-dimensional Poisson equation with steep solution: decomposition of the MSNC with the NN and FDM part (*gridNum1D* = 80).

**Table 2** Two-dimensional Poisson equation: RMSE with different *gridNum1D* and convergence order

| | *gridNum1D* | RMSE FDM | Order | RMSE MSNC | Order | Ratio |
|---|---|---|---|---|---|---|
| Smooth solution | 10 | 3.081058e-01 | | 1.278477e-02 | | 24.10 |
| | 20 | 6.370638e-02 | 2.42 | 1.983939e-03 | 3.22 | 32.11 |
| | 40 | 1.543939e-02 | 2.06 | 5.523248e-04 | 1.80 | 27.95 |
| | 80 | 3.854884e-03 | 2.00 | 1.439094e-04 | 1.92 | 26.79 |
| | 160 | 9.664398e-04 | 1.99 | 3.646306e-05 | 1.97 | 26.50 |
| | 320 | 2.421585e-04 | 2.00 | 9.057889e-06 | 2.01 | 26.73 |
| Steep Solution | 20 | 1.734401e-02 | | 8.081296e-04 | | 21.46 |
| | 40 | 5.138312e-03 | 1.69 | 1.099676e-04 | 3.67 | 46.73 |
| | 80 | 1.286862e-03 | 2.00 | 2.605703e-05 | 2.11 | 49.39 |
| | 160 | 3.228900e-04 | 1.99 | 6.575755e-06 | 1.98 | 49.10 |
| | 320 | 8.092320e-05 | 2.00 | 1.692923e-06 | 1.94 | 47.80 |

In summary, the above examples of two-dimensional PDEs demonstrates the MSNC 's notable advantages in both computational accuracy (exceeding a 20 times improvement) and cost (accompanied by a 16 times reduction in grid numbers) compared to the standard FDM.

## 3 Discussion

In this work, we propose a novel paradigm called MSNC for solving PDEs, with the consideration of the spectral bias of NNs and local approximation properties of numerical methods. Based on the idea of scale decomposition, NNs are employed for efficient capture of global scale, while numerical methods are utilized for detailed description of local scale. In addition, historical data is employed to train the NNs, and output layer weights of the trained NNs are released as free parameters in the new state of the PDEs.

Taking the hybrid of FDM and NNs as an example, the advantages of the MSNC have been validated in various numerical examples. (1) Higher computational accuracy. Under an equivalent grid number, the MSNC exhibits higher solution accuracy compared to the standard FDM, with an accuracy improvement of over 10 times for one-dimensional PDEs (in most situations) and over 20 times for two-dimensional PDEs. (2) Lower computational cost. With equivalent accuracy, the MSNC significantly reduces the required grid number compared to the standard FDM, with a 4 times reduction for one-dimensional PDEs, and a 16 times reduction for two-dimensional PDEs. (3) Stable convergence order of error. The MSNC demonstrates a stable RMSE convergence order with increasing grid numbers, facilitating a balanced consideration of computational accuracy and cost in practical applications. (4) Rigorous satisfaction of boundary conditions. The MSNC rigorously satisfies the given boundary conditions, better ensuring the solution to be well-posed. The MSNC demonstrates the feasibility and significant potential of hybrid of NNs and numerical methods. In our future work, we will further investigate the MSNC in more challenging examples, such as dealing with different types of boundary conditions, unsteady problems, and complex geometries.

## 4 Methods

### 4.1 The similarities between numerical methods and NNs-based PDEs solvers

Let us consider a (elliptic) PDE of the form

$$\sum_{i=1}^{d} \frac{\partial^2 u}{\partial x_i^{\,2}} = f(x), \quad x \in \Omega_d \tag{10}$$

where $i = 1,\ldots,d$ , $x = (x_1,\ldots,x_d) \in \Omega_d$ , and $d$ denotes the space dimension. Taking a one-dimensional problem as an example, the PDE (10) reduce to

$$\frac{\partial^2 u}{\partial x^2} = f(x), \quad x \in \Omega_1 \tag{11}$$

We select the 1D spatial domain $\Omega_1 = [a,b]$, and introduce a uniform grid defined by the points $x_i = x_0 + i\Delta x, \ i = 0,1,\ldots,n+1$ . Taking FDM as an example, the second derivative discretized by the second-order central difference is given by

$$\left.\frac{\partial^2 u}{\partial x^2}\right|_{x_i} = \frac{u(x_{i+1}) - 2u(x_i) + u(x_{i-1})}{\Delta x^2} + O(\Delta x^2) \tag{12}$$

It can be seen from Eq. (12) that, in the second-order central difference, the value of spatial derivative of $x_i$ is only depended on $x_i$ itself and the neighboring points $x_{i-1}$ and $x_{i+1}$ . It indicates that the information propagation in the FDM is local, or indirect, rather than global. For example, in the second-order central difference, the value at the boundary point $x_0$ directly influences the spatial derivative value at $x_1$ (due to $\left.\frac{\partial^2 u}{\partial x^2}\right|_{x_1} = \frac{u(x_2) - 2u(x_1) + u(x_0)}{\Delta x^2} + O(\Delta x^2)$ ), while indirectly influences the spatial derivative value at $x_2$ ( $\left.\frac{\partial^2 u}{\partial x^2}\right|_{x_2}$ is related to $u(x_1)$ ). Boundary information is propagated to the entire solution domain through this indirect manner.

Eq. (12) can be rewritten as (here we omit the truncation error term $O(\Delta x^2)$)

$$\left.\frac{\partial^2 u}{\partial x^2}\right|_{x_i} = \sum_{j=1}^{n} \alpha_j(x_i) u(x_j) \tag{13}$$

where $\alpha_{j-1}(x_i) = \alpha_{j+1}(x_i) = \frac{1}{\Delta x^2}, \alpha_j(x_i) = \frac{-2}{\Delta x^2}$ , with remaining terms taking the value of 0. This implies that the spatial derivative value at $x_i$ can be regarded as a linear combination of the $a_j(x_i)$ and $u(x_j)$ . In fact, $a_j(x_i)$ are the analytical derivative of the trial functions which serve as the basis functions of the approximate solution. In the FDM, trial functions are polynomial interpolation functions defined in the local solution domain. In summary, Eq. (13) indicates that the FDM approximates the solution by a series of trial functions defined in the local solution domain, with the weighting coefficients of the trial functions unknown.

Now, let us shift focus to the NNs-based PDEs solvers. First, consider the mapping relationship in the NNs. Inputs of a NN are passed layer by layer to the outputs through forward propagation. The output of a hidden layer can be represented as

$$H(\boldsymbol{x}) = \sigma(\boldsymbol{W}\boldsymbol{x} + \boldsymbol{b}) \tag{14}$$

where $\boldsymbol{x}$ denotes input vector, $\boldsymbol{W}$ denotes weight matrix, $\boldsymbol{b}$ denotes bias vector, and $\sigma$ denotes the activation function, which introduces non-linearity to the NN.

NNs with multiple hidden layers can be extended based on this. Output of a NN with $L$ hidden layers can be represented as

$$Y(\boldsymbol{x}) = \boldsymbol{W}^{(L+1)}(\boldsymbol{H}^{(L)}(\boldsymbol{H}^{(L-1)}\ldots(\boldsymbol{H}^{(1)}(\boldsymbol{x})))) + \boldsymbol{b}^{(L+1)} \tag{15}$$

where the superscript $(L)$ denotes the $L$-th hidden layer, and the superscript $(L+1)$ denotes the output layer, which is generally a linear layer.

Denote weights and biases in hidden layers by $\theta$, and omit the superscripts. Then the Eq. (15) reduce to

$$Y(\boldsymbol{x}) = \boldsymbol{W}H(\boldsymbol{x};\theta) + \boldsymbol{b} \tag{16}$$

Next, let us consider the NNs-based PDEs solvers. We begin with ELM-based PDEs solvers. In a ELM, unknown parameters are only output layer weights, while the parameters of other neurons are randomly generated. We define an ELM with a scope covering the entire solution domain, with input $x$ and output $u$, designed for solving the PDE defined by Eq. (11). In a ELM, the output layer biases are set to 0. Then referring to Eq. (16), the mapping relationship for this ELM can be expressed as

$$u(x) = \boldsymbol{W}H(x;\tilde{\theta}) \tag{17}$$

Once the ELM is defined, the $\theta$ in the hidden layers are fixed value. To distinguish from the optimizable $\theta$, we represent $\theta$ with the fixed values as $\tilde{\theta}$.

Employ AD for the ELM to compute spatial derivatives at uniform grid points, we obtain

$$\left.\frac{\partial^2 u}{\partial x^2}\right|_{xi} = \sum_{j=1}^{m} \frac{\partial h^2{}_j(x_i;\tilde{\theta})}{\partial x^2} w_j \tag{18}$$

Eq. (18) indicates that the spatial derivatives value at $x_i$ can be regarded as a linear combination of the derivatives of $h_j(x_i,\tilde{\theta})$ and the coefficients $w_j$. Comparing with Eq. (13), we observe that Eq. (18) and Eq. (13) share a similar form. The difference lies in the fact that $a_j(x_i)$ in Eq. (13) are local, while $\frac{\partial h^2{}_j(x_i;\tilde{\theta})}{\partial x^2}$ in Eq. (18) are global. Referring to Eq. (13), Eq. (18) indicates that the ELM approximates the solution by a series of trial functions defined in the global solution domain, with the weighting coefficients of the trial functions unknown.

Then, we consider the PINNs. Unlike ELM, all parameters of neurons in PINNs can be optimized. We define a PINN with a scope covering the entire solution domain, with input *x* and output *u*, designed for solving the PDE defined by Eq. (11). Referring to Eq. (16), the mapping relationship for this PINN can be expressed as

$$u(x) = \boldsymbol{W}H(x;\theta) + \boldsymbol{b} \tag{19}$$

where $\theta$ are optimizable in Eq. (19). Additionally, the output layer biases $\boldsymbol{b}$ in PINNs are generally not set to 0.

Employ AD for the PINN to compute spatial derivatives at uniform grid points. Now $\theta$ are also functions of the input *x*. However, to maintain formal consistency with ELM, we consider $\theta$ as an optimizable parameter in the expression, without explicitly writing out the derivative relationship between $\theta$ and *x*. Then the expression for the derivatives of Eq. (19) can then be represented as

$$\left.\frac{\partial^2 u}{\partial x^2}\right|_{xi} = \sum_{j=1}^{m} \frac{\partial h^2{}_j(x_i;\theta)}{\partial x^2} w_j \tag{20}$$

Comparing Eq. (18) and Eq. (20), we can observe that the PINN can be viewed as a linear combination of optimizable trial functions, whereas the ELM is a linear combination of randomly generated trial functions.

Comparing Eq. (13), (18), and (20) once again, we can conclude that the goal of the FDM is to solve the linear weighting coefficients for local trial functions, while the goals of the PINNs and ELM are to solve the linear weighting coefficients for global trial functions. From the perspective of the trial functions, NNs-based PDEs solvers and the FDM share similarities, both approximating the solution by a series of trial functions. However, the difference lies in the scopes and representation of these trial functions. In the FDM, trial functions are locally defined polynomial functions, whereas in NNs-based PDEs solvers, trial functions are globally defined neural networks. Additionally, although the analysis is conducted with a focus on the FDM, considering the similarities among the FDM and other numerical methods, such as the finite volume method (FVM) and the finite element method (FEM), these conclusions are applicable to other numerical methods as well.

### 4.2 Analysis of the generation of NNs

The NNs-based PDEs solvers introduce two main approaches: one is PINNs, where NNs parameters continuously optimize to approximate PDEs, and the other is ELM, where the output layer weights are the only unknowns, and parameters of other neurons are randomly generated. The following provides a brief analysis of these two approaches.

(I) PINNs for solving PDEs

Advantages: After optimizing the parameters, the NNs exhibit enhanced functions representation capabilities, making it more accurate for solving the targeted PDEs.

Disadvantages: The optimization of parameters costs a significant amount of time, leading to low solving efficiency.

(II) ELM for solving PDEs

Advantages: The generation of NNs parameters and solving PDEs come with low costs.

Disadvantages: Randomly generated parameters are often challenging to align with the inherent characteristics of the problem.

The analysis of the accuracy and efficiency of PINNs and ELM-based PDEs solvers suggests that their characteristics can complement each other. Therefore, is it possible to achieve a balance between accuracy and efficiency by combining the PINNs and ELM? Additionally, neural networks have the potential to leverage historical data, process data, and experimental data to reduce the computational costs of numerical simulations. Therefore, we adopt the following data-driven NNs generation strategy as described in Section 2.1.

The data-driven NNs generation strategy employed in this paper has the following advantages. In comparison to PINNs, the unknowns only include the output layer weights, eliminating the need for optimization of a large amount of parameters. In comparison to ELM-based PDEs solvers, the parameters of the NNs are more tailored to the problem being solved, reducing randomness.

### 4.3 Implementation process of the MSNC

To illustrate the specific implementation process of the MSNC, let us take the solution of the one-dimensional linear Helmholtz equation as an example.

Consider the one-dimensional linear Helmholtz equation with the first-type boundary conditions of the form

$$\begin{aligned} &\frac{\partial^2 u}{\partial x^2} - \lambda u = f(x), \quad x \in \Omega_1, \quad \Omega_1 = [a,b] \\ &u(a) = g_1, \quad u(b) = g_2 \end{aligned} \tag{21}$$

#### 4.3.1 The hybrid of NNs and FDM

Taking the FDM and the NNs as an example, we demonstrate the specific solving process of Equation (21) using the MSNC. We emphasize that this hybrid strategy is not limited to the FDM and can be easily extended to other numerical methods.

(I) PDEs discretization

We select the 1D spatial domain $\Omega_1 = [a,b]$, and introduce a uniform grid defined by the points $x_i = x_0 + i\Delta x,\ i = 0,1,\ldots,n+1$, where $x_0 = a, x_{n+1} = b$.

Decompose the variable $u$ in Eq. (21) into two parts, $u = \bar{u} + \hat{u}$, where $\bar{u}$ denotes the solution at the global scale, solved by the NN, and $\hat{u}$ denotes the solution at the local scale, solved by the FDM.

The NN has only the output layer weights as unknowns. Referring to Eq. (17), $\bar{u}$ can be expressed as:

$$\bar{u} = \bar{u}_0 + \sum_{j=1}^{m} w_j \bar{u}_j(x) \tag{22}$$

where $\bar{u}_0$ denotes the output layer bias of the NN. Due to only the DOFs of the output layer weights are released, $\bar{u}_0$ is a known constant. $w_j$ denotes the output layer weights to be solved, and $\bar{u}_j(x)$ denotes the outputs of the last hidden layer of the NN. Assuming the width of the last hidden layer in the NN is *m*, there are a total of *m* output layer weights to be solved.

For any interior point $x_i$, based on $u = \bar{u} + \hat{u}$, the second-order derivative term in Eq. (21) can be decomposed and expressed as:

$$\left.\frac{\partial^2 u}{\partial x^2}\right|_{xi} = \left.\frac{\partial^2 \bar{u}}{\partial x^2}\right|_{xi} + \left.\frac{\partial^2 \hat{u}}{\partial x^2}\right|_{xi} \tag{23}$$

Substituting Eq. (22), the derivatives of NN are computed using AD. The derivative of the bias term $\bar{u}_0$ is 0, and the second-order central difference is applied to the term $\frac{\partial^2 \hat{u}}{\partial x^2}$, we obtain

$$\left.\frac{\partial^2 u}{\partial x^2}\right|_{xi} = \sum_{j=1}^{m} w_j \frac{\partial^2 \bar{u}_j(x_i)}{\partial x^2} + \frac{\hat{u}(x_{i+1}) - 2\hat{u}(x_i) + \hat{u}(x_{i-1})}{\Delta x^2} \tag{24}$$

Combining Eq. (22) and Eq. (24), for any interior point $x_i$, Eq. (21) can be rewritten as:

$$[\sum_{j=1}^{m} w_j \frac{\partial^2 \bar{u}_j(x_i)}{\partial x^2} + \frac{\hat{u}(x_{i+1}) - 2\hat{u}(x_i) + \hat{u}(x_{i-1})}{\Delta x^2}] - \lambda[\bar{u}_0 + \sum_{j=1}^{m} w_j \bar{u}_j(x_i) + \hat{u}(x_i)] = f(x_i) \tag{25}$$

Moving the unknowns from Eq. (25) to the left side of the equal sign, and known terms to the right side, we obtain

$$\sum_{j=1}^{m}[\frac{\partial^2 \bar{u}_j(x_i)}{\partial x^2} - \lambda \bar{u}_j(x_i)]w_j + \frac{1}{\Delta x^2}\hat{u}(x_{i+1}) + (\frac{-2}{\Delta x^2} - \lambda)\hat{u}(x_i) + \frac{1}{\Delta x^2}\hat{u}(x_{i-1}) = f(x_i) + \lambda \bar{u}_0 \tag{26}$$

(II) Boundary conditions handling

For solving PDEs, the handling of boundary conditions needs to be considered. Here, we take the left boundary as an example. At the left boundary $x_0$,

$u(x_0) = \bar{u}(x_0) + \hat{u}(x_0)$, substituting into Eq. (22), we obtain

$$\hat{u}(x_0) = u(x_0) - (\bar{u}_0 + \sum_{j=1}^{m} w_j \bar{u}_j(x_0)) \tag{27}$$

At the point $x_1$, the second-order central difference of $\frac{\partial^2 \hat{u}}{\partial x^2}$ is related to $\hat{u}(x_0)$. Considering the discretization of the PDE at the point $x_1$, and substituting Eq. (27) into Eq. (26), we obtain

$$\sum_{j=1}^{m}[\frac{\partial^2 \bar{u}_j(x_1)}{\partial x^2} - \lambda \bar{u}_j(x_1)]w_j + \frac{1}{\Delta x^2}\hat{u}(x_2) + (\frac{-2}{\Delta x^2} - \lambda)\hat{u}(x_1) + \frac{1}{\Delta x^2}[u(x_0) - (\bar{u}_0 + \sum_{j=1}^{m} w_j \bar{u}_j(x_0))] = f(x_1) + \lambda \bar{u}_0 \tag{28}$$

$u(x_0)$ denotes the left boundary in Eq. (28), a known value defined in Eq. (21). Moving the unknowns from Eq. (28) to the left side of the equal sign, and known terms to the right side, we obtain

$$\sum_{j=1}^{m}[\frac{\partial^2 \bar{u}_j(x_1)}{\partial x^2} - \lambda \bar{u}_j(x_1) - \frac{\bar{u}_j(x_0)}{\Delta x^2}]w_j + \frac{1}{\Delta x^2}\hat{u}(x_2) + (\frac{-2}{\Delta x^2} - \lambda)\hat{u}(x_1) = f(x_1) + \lambda \bar{u}_0 - \frac{u(x_0) - \bar{u}_0}{\Delta x^2} \tag{29}$$

It can be seen that Eq. (29) can rigorously satisfy boundary conditions. Compared to applying different weights to PDEs and boundary conditions in the loss functions, rigorous satisfaction of boundary conditions can better ensure the solution to be well-posed.

(III) Algebraic system of equations

Applying Eq. (26) and (29) to the entire discrete system results in the form of an algebraic system of equations $AU = F$ as follows

$$A = \begin{bmatrix}
\frac{\partial^2 \bar{u}_1(x_1)}{\partial x^2} - \lambda \bar{u}_1(x_1) - \frac{\bar{u}_1(x_0)}{\Delta x^2} & \frac{\partial^2 \bar{u}_2(x_1)}{\partial x^2} - \lambda \bar{u}_2(x_1) - \frac{\bar{u}_2(x_0)}{\Delta x^2} & \cdots & \frac{\partial^2 \bar{u}_m(x_1)}{\partial x^2} - \lambda \bar{u}_m(x_1) - \frac{\bar{u}_m(x_0)}{\Delta x^2} & \frac{-2}{\Delta x^2} - \lambda & \frac{1}{\Delta x^2} & & & \\
\frac{\partial^2 \bar{u}_1(x_2)}{\partial x^2} - \lambda \bar{u}_1(x_2) & \frac{\partial^2 \bar{u}_2(x_2)}{\partial x^2} - \lambda \bar{u}_2(x_2) & \cdots & \frac{\partial^2 \bar{u}_m(x_2)}{\partial x^2} - \lambda \bar{u}_m(x_2) & \frac{1}{\Delta x^2} & \frac{-2}{\Delta x^2} - \lambda & \frac{1}{\Delta x^2} & & \\
\vdots & \vdots & \cdots & \vdots & & \ddots & \ddots & \ddots & \\
\frac{\partial^2 \bar{u}_1(x_{n-1})}{\partial x^2} - \lambda \bar{u}_1(x_{n-1}) & \frac{\partial^2 \bar{u}_2(x_{n-1})}{\partial x^2} - \lambda \bar{u}_2(x_{n-1}) & \cdots & \frac{\partial^2 \bar{u}_m(x_{n-1})}{\partial x^2} - \lambda \bar{u}_m(x_{n-1}) & & & \frac{1}{\Delta x^2} & \frac{-2}{\Delta x^2} - \lambda & \frac{1}{\Delta x^2} \\
\frac{\partial^2 \bar{u}_1(x_n)}{\partial x^2} - \lambda \bar{u}_1(x_n) - \frac{\bar{u}_1(x_{n+1})}{\Delta x^2} & \frac{\partial^2 \bar{u}_2(x_n)}{\partial x^2} - \lambda \bar{u}_2(x_n) - \frac{\bar{u}_2(x_{n+1})}{\Delta x^2} & \cdots & \frac{\partial^2 \bar{u}_m(x_n)}{\partial x^2} - \lambda \bar{u}_m(x_n) - \frac{\bar{u}_m(x_{n+1})}{\Delta x^2} & & & & \frac{1}{\Delta x^2} & \frac{-2}{\Delta x^2} - \lambda
\end{bmatrix} \tag{30}$$

$$U = \begin{bmatrix} w_1 \\ w_2 \\ \vdots \\ w_m \\ \hat{u}(x_1) \\ \hat{u}(x_2) \\ \vdots \\ \hat{u}(x_n) \end{bmatrix} \tag{31}$$

$$F = \begin{bmatrix} f(x_1) + \lambda \bar{u}_0 - \dfrac{u(x_0) - \bar{u}_0}{\Delta x^2} \\ f(x_2) + \lambda \bar{u}_0 \\ \vdots \\ f(x_{n-1}) + \lambda \bar{u}_0 \\ f(x_n) + \lambda \bar{u}_0 - \dfrac{u(x_{n+1}) - \bar{u}_{n+1}}{\Delta x^2} \end{bmatrix} \tag{32}$$

where $A$ denotes the *Jacobi* matrix with dimensions $n \times (m+n)$, $n$ is the number of interior grid points, and $m$ is the number of output layer weights in the NN. $U$ is the vector of unknowns for the algebraic system of equations, including the output layer weights of the NN and the local values at each grid point in FDM, with dimensions $(m+n) \times 1$.

After solving the algebraic system of equations, the solution for the entire solution domain is obtained by

$$u(x) = \bar{u}_0 + \sum_{j=1}^{m} w_j \bar{u}_j(x) + \hat{u}(x) \tag{33}$$

4.3.2 Algorithmic flow and implementation details

In addition, Section 4.3.1 only presented the solution process for linear PDEs. For nonlinear PDEs, there is an additional step of local linearization in the solution process. Taking the one-dimensional nonlinear Helmholtz equation as an example, the PDE can be expressed as:

$$\begin{aligned} &\frac{\partial^2 u}{\partial x^2} - \lambda u + \beta \sin(u) = f(x), \quad x \in \Omega_1, \quad \Omega_1 = [a,b] \\ &u(a) = g_1, \quad u(b) = g_2 \end{aligned} \tag{34}$$

Eq. (34) includes a nonlinear term $\beta \sin(u)$. The local linearization is utilized as follows. Considering an iterative process of solving the algebraic equation system, where $u^{(n-1)}$ denotes the solution at the previous iteration step (known in the current iteration step), and $u^{(n)}$ denotes the solution at the current iteration step (unknown). At the current iteration step, $\beta \sin(u)$ is linearized by substituting the value of $u$ from the previous iteration step, namely $\beta \sin(u^{(n-1)})$. The iteration is stopped when the error between two consecutive iterations of $u$ falls below a certain threshold.

The solution processes for linear and nonlinear PDEs using the MSNC is summarized in Algorithm 1.

The code implementation of the MSNC in this paper is based on Python 3. The NNs implementation relies on the open-source deep learning library PyTorch[31], while the storage and solution of algebraic equation systems use the scientific computing library SciPy[32].

## 4.4 NNs model Training

Taking Eq. (21) as an example, the NN in this paper is trained based on historical data to establish the mapping between the spatial coordinates $x$ of the solution domain and the PDEs solution $u$. Simultaneously, to distinguish historical data for different state parameters and achieve generalization of state parameters, these state parameters $c$ need to be included as inputs to the NN. In summary, the NN takes spatial coordinates $x$ (the number of spatial coordinates depends on the dimensionality of the PDEs) and state parameters c as inputs, and outputs the solution u of the PDEs.

Using the residual neural network as the basic model architecture, it improves upon the framework of fully connected neural networks. The basic building block is changed from a hidden layer to a residual block, as shown in Figure 4(a). The output of a residual block is the linear sum of the input and output of a hidden layer. Residual neural networks exhibit better training convergence efficiency compared to fully connected neural networks. Based on the residual neural network, the NN model established in this paper is illustrated in Figure 4(b). The model takes the value of spatial coordinates and state parameters at a single point as inputs, and outputs the corresponding solution of the PDEs at that single point.

Latin Hypercube Sampling (LHS) is employed to select historical data state parameters, and then, through numerical solutions or analytical solutions, construct data samples for the selected state parameters. After generating the data, the dataset, consisting of spatial coordinates, state parameters, and PDEs solutions, is split into training and testing sets with an 80% and 20% ratio, respectively.

The loss function is set as the mean squared error (MSE) between the predicted values and the label values. In the actual training process of the NN model, data is

**Algorithm 1: Pseudo-algorithm of the proposed methodology**

// Data preparation

1: collect historical data with multiple values of state parameters $\boldsymbol{c}$

// NN training

2: train NN model as $u = u(\boldsymbol{x}, \boldsymbol{c})$

// NN transfer

3: transfer NN model by releasing DOFs of output layer weights

// PDEs discretization

4: split terms in PDEs related to $u$ with expression $u = \bar{u} + \hat{u}$

5: discretize derivative in global scale (related to $\bar{u}$ ) with AD, and derivative in local scale (related to $\hat{u}$ ) with the FDM

6: deal with the boundary conditions

// Algebraic equations solving

7: **if** PDEs is linear **then**

8:     solve linear system $AU = F$

9: **end**

10: **if** PDEs is nonlinear **then**

11:     initialize $U$: $U \leftarrow U^{(0)}$

12:     while $abs(U^{(n)} - U^{(n-1)}) < \epsilon$

13:         update U: $U^{(n-1)} \leftarrow U^{(n)}$

14:         linearize nonlinear term in PDEs with $U^{(n-1)}$

15:         solve linear system $AU^{(n)} = F$

16: **end**

// Results processing

17: get solution of PDEs with expression $u = \bar{u}_0 + \sum_{j=1}^{m} w_j \bar{u}_j + \hat{u}$

read in batches, so the loss function can be expressed as

$$LOSS_{MSE}(u) = \frac{1}{N}\sum_{i=1}^{N}(u_{pred} - u_{ref})^2 \tag{35}$$

a)

b)

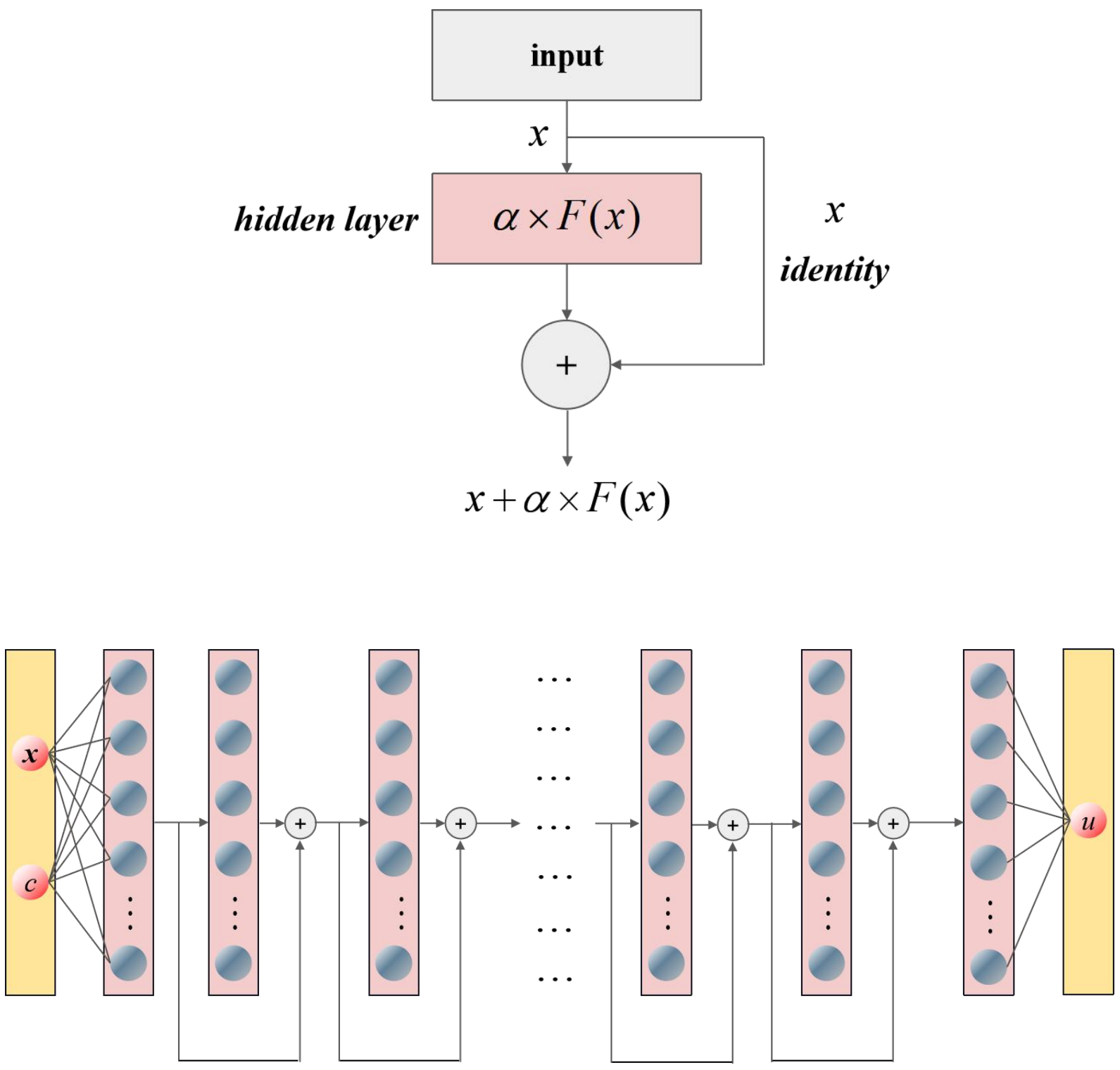

Figure 4: **a)** Schematic diagram of a residual block. **b)** Schematic diagram of the NN model established in this paper.

where $N$ denotes the size of a batch, $u_{pred}$ denotes the the predicted values, and $u_{ref}$ denotes the label values.

In addition, the sine function is employed as the activation function, and AdamW is used as the optimizer for model training.

For the numerical examples in this paper, two state parameters $\alpha_1$ and $\alpha_2$ in the source term $f(x)$ are sampled by LHS to obtain the historical data. The sampling space for the state parameters $\alpha_1$ and $\alpha_2$ chosen as $[2.9, 3.1] \times [1.9, 2.1]$ in one-dimensional linear Helmholtz equation, one-dimensional Poisson equation, and one-dimensional nonlinear Helmholtz equation, $[0.9, 1.1] \times [3.9, 4.1]$ in two-dimensional Poisson equation with smooth solution, and $[1.9, 2.1] \times [1.9, 2.1]$ in two-dimensional Poisson equation with steep solution.

After obtaining the historical data, the NNs are trained using the architecture and parameters settings introduced above, with a width of 10 and a depth of 6 for one-dimensional linear Helmholtz equation, a width of 20 and a depth of 11 for one-dimensional Poisson equation, a width of 20 and a depth of 4 for two-dimensional , a width of 40 and a depth of 3 for two-dimensional Poisson equation with smooth solution, and a width of 40 and a depth of 9 for two-dimensional Poisson equation with steep solution.

We emphasize that obtaining historical data in this paper is just a necessary step to validate the performance of the proposed method. It does not imply that in practical applications, users must perform multiple numerical simulations to obtain data and then train a NN. In fact, the envisioned practical applications in this paper is as follows. Users have already accumulated a historical database, and then a generalized NN is established based on this historical database for the full utilization of data resources. When solving new problems, users just need to apply the MSNC proposed in this paper based on the established NN.